\documentclass[
twocolumn,
longbibliography,
showpacs,                     %  Add 'showpacs' option to make PACS codes appear
showkeys,                  %  Add 'showkeys' option to make keywords appear
secnumarabic,
amssymb,
nobibnotes,
aps,
pre]{revtex4-1}

\usepackage{graphicx}
\usepackage{placeins}
\usepackage{latexsym}
\usepackage{amsmath}
\usepackage{verbatim}

\begin{document}

\title{The surprising convergence of the 
Monte Carlo renormalization group for the d=3 Ising Model}

\author{Dorit Ron}
%\affiliation{Faculty of Mathematics and Computer Science, The Weizmann Institute of Science, Rehovot 76100, Israel}
  \email[]{dorit.ron@weizmann.ac.il}

\author{Achi Brandt}
\email[]{achi.brandt@weizmann.ac.il}

\affiliation{Faculty of Mathematics and Computer Science, The Weizmann Institute of Science, Rehovot 76100, Israel }

\author{Robert H. Swendsen}
\email[]{swendsen@cmu.edu}
\affiliation{Department of Physics, Carnegie Mellon University, Pittsburgh, Pennsylvania, 15213, USA}
%\thanks{}
%\altaffiliation{}


\date{\today}

\begin{abstract}
We present a surprisingly simple approach
to high-accuracy calculations
of critical properties of
the three-dimensional Ising model.
The method uses a modified
block-spin transformation
with a tunable parameter
to improve convergence
in Monte Carlo renormalization group.
The block-spin parameter must be tuned
differently for different exponents
to produce optimal convergence.
\end{abstract}

\maketitle

\section{Introduction}\label{section: Introduction}

The Monte Carlo renormalization group (MCRG) method
is a systematic procedure for computing critical properties
of lattice spin models\cite{Ma_MCRG_1976,RHS_MCRG_1979}.
It has been shown to be both flexible and effective
in the calculation of critical exponents,
critical temperatures,
and renormalized couplings constants\cite{RHS_MCRG_optimize_d=2,RHS_MCRG_optimize_d=3,Brandt_Ron_2001,Ron_RHS_2001,Ron_RHS_2002,Ron_RHS_Brandt_2002,Ron_RHS_Brandt_2005}.
A particularly interesting application of MCRG is the three-dimensional
Ising model\cite{Bloete_RHS_3d_Ising_1979,Pawley_MCRG_3d_Ising_1984}.
This model has proven to be one of the most difficult to obtain accurate estimates for,
because the approach to the fixed point is so slow.
Attempts have been made to bring the fixed point closer to the
 nearest-neighbor model\cite{RHS_MCRG_optimize},
but these have been controversial\cite{Fisher_Randeria},
and have not resulted in improved results.

The most encouraging result has been that of
Bl\"ote et al.\cite{Bloete_Heringa_Hoogland_Meyer_Smit_1996}
who used a three-parameter approximation to the fixed point,
along with a modified majority rule for the RG transformation.
We have discovered a particularly simple modification 
of this calculation,
which simulates  the nearest-neighbor critical point 
and optimizes the RG transformation for the even and odd exponents
separately.

In the following sections,
we recall the MCRG method,
illustrate the slow convergence with the majority rule,
and introduce  a  tunable RG transformation
in  section \ref{section_tuning}\cite{Bloete_Heringa_Hoogland_Meyer_Smit_1996}.
The improved convergence of the tuned RG transformation
is  then demonstrated in sections \ref{section_largest_even} 
through \ref{section_correlation_functions}.
Finally, we present our conclusions and 
discuss future work.

\section{MCRG computations}
\label{section_MCRG}

We consider the three-dimensional Ising model on a simple cubic lattice,
of size $N \times N \times N$.
The Hamiltonian is given by
\begin{equation}
H = K \sum_{\langle j,k \rangle} \sigma_j    \sigma_k    ,
\end{equation}
where
$\sigma_j = \pm 1$,
and
the sum is over all nearest-neighbor pairs.
The dimensionless coupling constant
$K$ includes the inverse temperature
$\beta=1/k_B T$,
so as to make the Boltzmann factor $e^H$.

We used Wolff algorithm\cite{Wolff_1989},
to simulate the model at an inverse temperature of
 $K_c=0.2216544$\cite{Talapo_Blote}.
 The renormalized configurations were obtained from these sets.
For each configuration, the lattice was divided up into   cubes, each containing eight sites,
so that the scaling factor $b=2$.
 We will denote this block of spins, as well as the renormalized spins associated with them
 by $\ell$.
 A value of plus or minus one was assigned to each renormalized spin to represent the  original spins in each cube.
We used the 
 ran2 random number generator from Numerical Recipes\cite{Numerical_Recipes}.
The lengths of the simulations we used are given in Table \ref{table: YT1 data}.

\begin{table}[h]
\caption{Data for the  $y_{T1}$ simulations.
%The simulations were performed at  $K_c=0.2216544$\cite{Talapo_Blote}.
}
\begin{center}
\begin{tabular}{|c|c|c|c|c|c|}
\hline
& $256^3$ & $128^3$ & $64^3$ & $32^3$ & $16^3$  \\
\hline
 $\#_{\textrm{sites}}/N^3$  & $2.2\times 10^5$ & $2.2\times 10^5$ & $2.2\times 10^5$ & $4\times 10^7$ & $4\times 10^8$ \\
\hline
 $\#_{\textrm{Wolff}}/N^3$  & 22 & 91 & 420 & $3.2 \times 10^4$ & $1.2 \times 10^6$ \\
\hline
$\Delta \#_{\textrm{Wolff}}$  & 164 & 87 & 50 & 26 &  15 \\
%$N_{\textrm{Wolff}}(\simeq N^3)$ between data  & 164 & 87 & 50 & 26 &  15 \\
\hline
 Cluster   & $1.0 \times 10^5$ & $2.4  \times 10^4$  & $5.2  \times 10^3$  & $1.2  \times 10^3$ & $2.7 \times 10^2$  \\
%Average cluster   & 102,080 & 24,166 & 5,268 & 1,239 & 273 \\
%Size of average cluster   & 102,080(17,102) & 24,166(3,882) & 5,268(528) & 1,239(49) & 273(14)  \\
\hline
\end{tabular}
\end{center}
\label{table: YT1 data}
\end{table}

The renormalized configurations can  be described by the set of
(unknown)
renormalized coupling constants,
$ K_{\alpha}^{(n)}$.
The subscript $\alpha$ denotes the type of coupling
(nearest-neighbor, next-nearest-neighbor, four-spin, etc.).
The nearest-neighbor coupling constant $K$ defined earlier,
is also  denoted by
$K_{nn}^{(0)}$.
All other coupling constants at level $n=0$  vanish.

To determine the critical exponents,
we then wish to calculate
the matrix of derivatives of the couplings at level $n+1$
with respect to the couplings at level $n$.
\begin{equation}\label{T-matrix}
T_{\alpha,\beta}^{(n+1,n)}
=
\frac{
\partial  K_{\alpha}^{(n+1)}
}{
\partial  K_{\beta}^{(n)}
}
\end{equation}
This matrix of derivatives is
 then given by the solution of  the equation
\begin{equation}
\frac{ \partial \langle  S_{\gamma}^{(n+1)} \rangle  }
{\partial  K_{\beta}^{(n)} }
=
\sum_{\alpha}
\frac{ \partial \langle  S_{\gamma}^{(n+1)} \rangle  }
{\partial  K_{\alpha}^{(n+1)} }
\frac
{\partial  K_{\alpha}^{(n+1)} }
{\partial  K_{\beta}^{(n)} }
\end{equation}
where
\begin{equation}
\frac{ \partial \langle  S_{\gamma}^{(n+1)} \rangle  }
{\partial  K_{\beta}^{(n)} }
=
\left\langle  S_{\gamma}^{(n+1)}  S_{\beta}^{(n)}  \right\rangle
-
\left\langle  S_{\gamma}^{(n+1)}  \right\rangle
\left\langle  S_{\beta}^{(n)}  \right\rangle   ,
\end{equation}
and
\begin{equation}
\frac{ \partial \langle  S_{\gamma}^{(n+1)} \rangle  }
{\partial  K_{\alpha}^{(n+1)} }
=
\left\langle  S_{\gamma}^{(n+1)}  S_{\alpha}^{(n+1)}  \right\rangle
-
\left\langle  S_{\gamma}^{(n+1)}   \right\rangle
\left\langle S_{\alpha}^{(n+1)}  \right\rangle   .
\end{equation}

For our calculations  we have included $N_e=30$ even and $N_o=20$ odd interactions.
%We have followed\cite{Baillie_PhysRevD_46_2480},   THE RIGHT PAPER IS BELOW:
We have followed\cite{Baillie_Gupta_Hawick_Pawley_1992},
 who calculated all $53$ even and $46$ odd interactions that fit in either a $3 \times 3$ square or a $2 \times 2 \times 2$ cube of spins, and used their $34$ first even operators (excluding the $25^{th}$,$29^{th}$,$31^{st}$ and the $32^{nd}$) and their first $20$ odd operations.
%
%Preliminary calculations using up to ?? even and ?? odd operators did not reveal any differences.
%
The eigenvalues of the $T$-matrix
in Eq.~(\ref{T-matrix})
are found separately for the even and odd operators.
The critical exponents are then obtained from the usual equations.

\section{The slow convergence of the majority rule}
\label{section_slow}

The usual majority rule performs very well for the two-dimensional Ising model,
which converges to the fixed point values of the exponents by the second iteration
of the RG transformation\cite{RHS_MCRG_1979}.
For the three-dimensional Ising model, however,
convergence is very slow,
as shown in Table \ref{table: YT1_majority}.
Reading across at each level $n$ of RG iterations,
the values of the approximations for
$y_{T1}$ are quite consistent.
There is no problem with finite-size effects on the RG trajectories,
indicating that the range of the renormalized Hamiltonian is limited.
However,
 even after five iterations of the renormalization group,
the value of $y_{T1}$
does not seem to have converged.
Anticipating our final result of
$y_{T1} \approx 1.591$,
Table  \ref{table: YT1_majority}
is very far from convergence.

\begin{table}[htb]
\caption{The eigenvalue exponent $y_{T1}$ for majority transformation which is equivalent to using a large $w$ in Eq.~(\ref{RG_transform}).
%The simulations were performed at  $K_c=0.2216544$.
}
\begin{center}
\begin{tabular}{|c|c|c|c|c|c|}
\hline
$n$  & $N_{e}$ & $256^3$ & $128^3$ & $64^3$ & $32^3$  \\
%$n$  & $N_{ops}$ & $L=8$ & $L=7$ & $L=6$ & $L=5$   \\
\hline
1 & Ê10 & 1.4189(7) & 1.4202(5) & 1.4203(2) & 1.4206(1)  \\
 & Ê20 & 1.4230(7) & 1.4241(5) & 1.4240(2) & 1.4241(1)  \\
 & Ê30 & 1.4224(7) & 1.4243(5) & 1.4237(2) & 1.4238(1)  \\
% & Ê40 & 1.4222(8) & 1.4238(5) & 1.4232(2) & 1.4234(1)  \\
% & Ê50 & 1.4224(8) & 1.4238(5) & 1.4234(3) & 1.4233(1)  \\
\hline
2 & Ê10 & 1.5093(6) & 1.5106(3) & 1.5108(1) & 1.5120(1)     \\
 & Ê20 & 1.5076(6) & 1.5088(3) & 1.5086(1) & 1.5084(1)     \\
 & Ê30 & 1.5072(6) & 1.5084(3) & 1.5082(1) & 1.5076(1)     \\
% & Ê40 & 1.5071(6) & 1.5084(3) & 1.5082(2) & 1.5080(1)     \\
% & Ê50 & 1.5069(6) & 1.5083(4) & 1.5080(1) & 1.5080(1)     \\
\hline
3 & Ê10 & 1.5521(5) & 1.5534(5) & 1.5544(2) &      \\
 & Ê20 & 1.5508(5) & 1.5515(5) & 1.5507(2) &      \\
 & Ê30 & 1.5504(5) & 1.5512(5) & 1.5501(2) &      \\
 %& Ê40 & 1.5504(5) & 1.5513(5) & 1.5506(2) &      \\
% & Ê50 & 1.5504(5) & 1.5512(5) & 1.5506(3) &      \\
\hline
4 & Ê10 & 1.5733(7) & 1.5745(5) &  &      \\
 & Ê20 & 1.5721(7) & 1.5711(5) &  &      \\
 & Ê30 & 1.5718(7) & 1.5702(5) &  &      \\
% & Ê40 & 1.5718(7) & 1.5702(6) &  &      \\
% & Ê50 & 1.5719(8) & 1.5706(9) &  &      \\
\hline
5 & Ê10 & 1.5825(11) &  &  &      \\
  & Ê20 & 1.5797(11) &  &  &      \\
 & Ê30 & 1.5788(11) &  &  &      \\
% & Ê40 & 1.5796(12) &  &  &      \\
 %& Ê50 & 1.5801(14) &  &  &      \\
\hline
\end{tabular}
\end{center}
\label{table: YT1_majority}
\end{table}

\section{Tunable block-spin transformation}
\label{section_tuning}

 Instead of using the usual majority rule, the renormalized spin
  was assigned a value according
   to the following probability\cite{Bloete_Heringa_Hoogland_Meyer_Smit_1996}.

\begin{equation}\label{RG_transform}
P( \sigma'_{\ell} ) = \frac{
\exp ( w \, \sigma'_{\ell} \sum_{j \in \ell} \sigma_j  )
}{
\exp ( w \sum_{j \in \ell}  \sigma_j  )  +  \exp ( - w \sum_{j \in \ell}  \sigma_j )
}
\end{equation}
For $w = \infty$, this becomes identical to the majority rule.

A special feature of the present calculation is that the RG transformation
in Eq.~(\ref{RG_transform})
was optimized separately for the even and odd operators.
The determination of the optimal value of the parameter  $w$
was also done much more carefully than in earlier work.
The
value of $w$ was adjusted
so that the
 largest eigenvalue
(for the even and odd operators separately)
was nearly constant for $n>1$.

\section{The largest even eigenvalue exponent, $y_{T1}$}
\label{section_largest_even}

The results for the largest even eigenvalue
are given
 in Table \ref{table: YT1}.
By contrast to the slow convergence seen in
Table \ref{table: YT1_majority},
the convergence
is striking.
For $n=1$,  the majority rule has $y_{T1}\approx 1.422$,
changes to
$y_{T1}\approx 1.507$ for $n=2$,
and increases to
$y_{T1}\approx 1.579$
for $n=5$.
The tuned RG transformation
 in Table \ref{table: YT1}
starts with
$y_{T1}\approx 1.593$ for $n=1$,
moves to
$y_{T1}\approx 1.592$ for $n=2$,
and stays at
$y_{T1}\approx 1.591$
for $n=3,4, \textrm{and } 5$.
Our best estimate for the largest even eigenvalue is
$y_{T1}= 1.591(1)$.

\begin{table}[htb]
\caption{The eigenvalue exponent $y_{T1}$.
The parameter $w=0.4314$.
%The simulations were performed at  $K_c=0.2216544$.
}
\begin{center}
\begin{tabular}{|c|c|c|c|c|c|c|}
\hline
$n$  & $N_{e}$ & $256^3$ & $128^3$ & $64^3$ & $32^3$ & $16^3$  \\
%$n$  & $N_{ops}$ & $L=8$ & $L=7$ & $L=6$ & $L=5$ & $L=4$  \\
\hline
1 & 10 & 1.5870(6) & 1.5864(2) & 1.5865(1) & 1.5872(1) & 1.58835(4) \\
 & Ê20 & 1.5923(7) & 1.5914(3) & 1.5916(1) & 1.5920(1) & 1.59266(5) \\
 & Ê30 & 1.5930(8) & 1.5924(3) & 1.5922(2) & 1.5927(2) & 1.59308(5) \\
\hline
2 &Ê10 & 1.5908(5) & 1.5903(2) & 1.5907(1) & 1.5920(1) &    \\
 & Ê20 & 1.5917(5) & 1.5912(2) & 1.5915(1) & 1.5923(1) &    \\
 & Ê30 & 1.5919(5) & 1.5914(2) & 1.5916(1) & 1.5921(1) &    \\
\hline
3 & Ê10 & 1.5910(5) & 1.5904(2) & 1.5922(1) &  &    \\
 & Ê20 & 1.5912(5) & 1.5905(2) & 1.5920(1)  &  &    \\
 & Ê30 & 1.5912(5) & 1.5905(2) & 1.5918(1) &  &    \\
\hline
4 & Ê10 & 1.5906(6) & 1.5918(3) &  &  &    \\
 & Ê20 & 1.5906(6) & 1.5914(3) &  &  &    \\
 & Ê30 & 1.5906(6) & 1.5912(3) &  &  &    \\
\hline
5 & Ê10 & 1.5911(6) &  &  &  &    \\
 & Ê20 &  1.5909(6) &  &  &  &    \\
 & Ê30 &  1.5908(7) &  &  &  &    \\
\hline
\end{tabular}
\end{center}
\label{table: YT1}
\end{table}

\section{The largest odd eigenvalue exponent, $y_{H1}$}
\label{section_largest_odd}

Table \ref{table: YH1_0.555}
shows the convergence of $y_{H1}$
for the tuned renormalization group.
Reading along the rows,
we see that there is virtually no effect of the size of the lattice
on the estimated values of $y_{H1}$.
Only for a renormalized lattice of $4 \times 4 \times 4$
can a decrease in the value of $y_{H1}$
of about $0.00014$ be seen.
Neither is there a noticeable dependence
on the number of operators
for a given number of RG iterations $n$.

The first iteration of the renormalization group
($n=1$)
gives an estimate of about $y_{H1}\approx 2.5086$.
For $n=2$,
it has dropped slightly to
$y_{H1}\approx 2.48507$,
and for
$n=3, 4, \textrm{and } 5$,
it is
$y_{H1}\approx 2.4829$.
Our best estimate is
$y_{H1}= 2.4829(2)$.

\begin{table}[htb]
%\tiny
\caption{The eigenvalue exponent $y_{H1}$.
The parameter $w=0.555$.
%The simulations were performed at  $K_c=0.2216544$.
}
\begin{center}
\begin{tabular}{|c|c|c|c|c|c|c|}
\hline
$n$  & $N_{o}$ & $256^3$ & $128^3$ & $64^3$ & $32^3$ & $16^3$  \\
\hline						
1 &	5	&  2.50830(8)   &  2.50831(3)   &   2.50829(1) &  2.50828(1) &   2.50823(1) \\	
& 10	&  2.50853(14)  &  2.50869(4)	&   2.50860(2) &  2.50860(2) & 	 2.50856(1) \\		
&	 15	&  2.50844(15)  &  2.50871(5)	&   2.50859(2) &  2.50861(2) &   2.50859(1) \\		
&	20	&  2.50843(15)  &  2.50871(5)	&   2.50860(2) &  2.50862(2) &   2.50860(1) \\
\hline						
2 & 5	&  2.48503(2) &  2.48504(1)	&   2.48503(1) &  2.48503(2) &   2.48481(1) \\	
&	 10	&  2.48507(3) &  2.48506(1)	&   2.48505(1) &  2.48506(2) &   2.48498(1) \\	
&	 15	&  2.48508(3) &  2.48507(1)	&   2.48506(1) &  2.48507(2) &   2.48493(2) \\	
&	20	&  2.48508(3) &  2.48507(1)	&   2.48506(1) &  2.48507(2) &   2.48494(2) \\
\hline					
3 & 5	&  2.48285(3) &  2.48287(2)	&  2.48286(1) &   2.48267(4) & \\
&	 10	&  2.48284(3) &  2.48287(2)	&  2.48288(1) &   2.48279(4)& \\			
&	 15	&  2.48285(3) &  2.48288(2)	&  2.48288(1) &   2.48274(4)& \\		
&	20	&  2.48285(3) &  2.48288(2)	&  2.48288(1) &   2.48274(4)& \\
\hline						
4 &	5	&  2.48300(5)	 &  2.48293(4)	&  2.48279(3) & &	\\		
&	 10	&  2.48299(5)	 &  2.48295(4)	&  2.48290(3) & &	\\		
&	 15	&  2.48300(5)	 &  2.48295(4)	&  2.48283(3) & &	\\	
&	20	&  2.48300(5)	 &  2.48294(4)	&  2.48282(3) & &	\\
\hline						
5  & 5	&  2.48271(12)	&  2.48250(9) & & &			\\		
&	 10	&  2.48272(12)	&  2.48259(9) & & &			\\		
&	 15	&  2.48273(12)	&  2.48251(9) & & &	\\	
&	20	&  2.48274(12)	&  2.48251(9) & & &  \\
\hline						
6  & 5	&  2.48223(23)	&  & & &			\\		
&	 10	&  2.48234(24)	&  & & &			\\		
&	 15	&  2.48234(25)	&  & & &	\\	
&	20	&  2.48231(25)	&  & & &  \\
\hline
\end{tabular}
\end{center}
\label{table: YH1_0.555}
\end{table}

\section{ The second-largest even eigenvalue exponent, $y_{T2}$}
\label{section_second_largest_even}

Table \ref{table: YT2}
shows that estimates for the second even eigenvalue
as a function of the number of RG iterations, $n$,
and the size of the renormalized lattices.
This eigenvalue is negative (``irrelevant''),
and controls the leading corrections to scaling.

The second largest eigenvalues
are naturally not as accurately determined as the largest.
We need about 20 operators to see the asymptotic behavior.
There is a slight trend for the values of
$y_{T2}$
to increase in magnitude with an increasing number of RG iterations,
suggesting that the asymptotic eigenvalue exponent
is actually larger.
Note that
similar slow convergence for the calculation of $y_{T2}$ was already observed and reported by Baillie et al.\cite{Baillie_Gupta_Hawick_Pawley_1992}.
Perhaps we can estimate
$y_{T2}= -0.75(5)$
from the tables,
but that might be overly optimistic.

The correction-to-scaling exponent
is given by the ratio of
$\omega=-y_{T2}/y_{T1}$,
so that we would estimate
$\omega= 0.75/1.591=0.47(3)$.

\begin{table}[htb]
\caption{The eigenvalue exponent $y_{T2}$.
The parameter $w=0.4314$, the tuned parameter for $y_{T1}$.
The simulations were the same as for $y_{T1}$,
which are given in Table \ref{table: YT1}.}
\begin{center}
\begin{tabular}{|c|c|c|c|c|c|c|}
\hline
$n$  & $N_{e}$ & $256^3$ & $128^3$ & $64^3$ & $32^3$ & $16^3$  \\
%$n$  & $N_{ops}$ & $L=8$ & $L=7$ & $L=6$ & $L=5$ & $L=4$  \\
\hline
1 & 10 &  -0.60(2) & -0.548(6) &  -0.545(2) &  -0.546(3) &  -0.5300(10) \\
 & Ê20 &  -0.70(2) & -0.622(9) &  -0.616(3) &  -0.622(4) &  -0.6081(12)\\
 & Ê30 &  -0.67(2) & -0.644(12) & -0.626(3) &  -0.638(4) &  -0.6186(12)\\
\hline
2 &Ê10 &  -0.64(2) & -0.608(5) &  -0.611(2) &  -0.607(2) &    \\
 & Ê20 &  -0.70(2) & -0.664(7) &  -0.668(2) &  -0.668(3) &    \\
 & Ê30 &  -0.71(2) & -0.698(10) & -0.691(3) &  -0.688(4) &    \\
\hline
3 & Ê10 &  -0.66(2) & -0.635(5) &  -0.634(2) &  &    \\
 & Ê20 &   -0.71(2) & -0.691(8) &  -0.696(3)  &  &    \\
 & Ê30 &   -0.72(2) & -0.722(10) & -0.725(4) &  &    \\
\hline
4 & Ê10 &  -0.67(1) &  -0.658(6) &  &  &    \\
 & Ê20 &   -0.74(2) &  -0.726(9) &  &  &    \\
 & Ê30 &   -0.73(2) &  -0.753(10) &  &  &    \\
\hline
5 & Ê10 &  -0.73(2) &  &  &  &    \\
 & Ê20 &   -0.77(2) &  &  &  &    \\
 & Ê30 &   -0.76(2) &  &  &  &    \\
\hline
\end{tabular}
\end{center}
\label{table: YT2}
\end{table}

\section{The second-largest odd eigenvalue exponent, $y_{H2}$}
\label{section_second_largest_odd}

The second-largest odd eigenvalue exponent,
like its counterpart in the two-dimensional Ising model,
is positive (``relevant'').
It has smaller statistical errors than its even counterpart,
but it also shows a slightly slower convergence.
As shown in
Table \ref{table: YH2b},
the first iteration of the RG transformation
($n=1$)
is $0.287(6)$,
and therefore quite far from the best estimate.
By $n=3$ and greater,
the value of the second odd eigenvalues
 has converged to about
$y_{H2}=0.403(4)$.

\begin{table}[htb]
\caption{The eigenvalue exponent $y_{H2}$.
The parameter $w=0.555$, the tuned parameter for $y_{H1}$.
The simulations were the same as for $y_{H1}$,
which are given in Table \ref{table: YH1_0.555}.
}
\begin{center}
\begin{tabular}{|c|c|c|c|c|c|c|}
\hline
$n$  & $N_{o}$ & $256^3$ & $128^3$ & $64^3$ & $32^3$ & $16^3$  \\
\hline	
1 &	5 &  0.232(4) &  0.236(1) &  0.2409(6) &  0.2469(7) &  0.2561(2) \\	
& 10  &  0.287(5) &  0.287(2) &  0.2888(7) &  0.2910(8) &  0.2935(3) \\		
& 15  &  0.287(6) &  0.290(2) &  0.2911(8) &  0.2931(9) &  0.2957(3) \\		
& 20  &  0.287(6) &  0.297(2) &  0.2972(8) &  0.2992(9) &  0.3012(3) \\
\hline						
2 & 5	&  0.306(3) &  0.307(1) &  0.3152(4)  &  0.3284(5) &  0.3405(2) \\	
&	 10	&  0.362(4) &  0.360(2) &  0.3635(5)  &  0.3689(6) &  0.3677(2) \\
&	 15	&  0.366(4) &  0.365(2) &  0.3680(5)  &  0.3728(6) &  0.3744(2) \\	
&	20	&  0.371(5) &  0.372(2) &  0.3735(5)  &  0.3777(6) &  0.3770(2) \\
\hline					
3 & 5	&  0.318(3) &  0.331(1) &  0.3455(5) &   0.3586(5) & \\
&	 10	&  0.379(3) &  0.384(1) &  0.3900(6) &   0.3871(6) & \\		
&	 15	&  0.384(3) &  0.390(1) &  0.3949(6) &   0.3949(6) & \\		
&	20	&  0.392(3) &  0.396(2) &  0.4000(6) &   0.3976(7) & \\
\hline						
4 &	5	&  0.335(3)	&  0.353(1) &  0.3638(4) & &	\\		
&	 10	&  0.390(3)	&  0.400(1) &  0.3935(4) & &	\\		
&	 15	&  0.396(3)	&  0.406(1) &  0.4024(4) & &	\\	
&	20	&  0.402(3)	&  0.411(1) &  0.4053(5) & &	\\
\hline
5  & 5	&  0.349(3)	&  0.366(1) & & &			\\		
&	 10	&  0.397(3)	&  0.397(1) & & &			\\		
&	 15	&  0.403(3)	&  0.407(1) & & &	\\	
&	20	&  0.410(3)	&  0.410(1) & & &  \\
\hline						
6  & 5	&  0.364(3)	&  & & &			\\		
&	 10	&  0.395(3)	&  & & &			\\		
&	 15	&  0.406(3)	&  & & &	\\	
&	20	&  0.409(3)	&  & & &  \\
\hline
\end{tabular}
\end{center}
\label{table: YH2b}
\end{table}

\section{Convergence of the correlation functions}
\label{section_correlation_functions}

Note that Bl\"ote et al. were able to achieve improved
 convergence with $w=0.4$ and a different Hamiltonian,
 which appeared to be closer to the fixed point\cite{Bloete_Heringa_Hoogland_Meyer_Smit_1996}.
 %\cite{Bloete_Luijten_Heringa_1995}.
Our results show that the improved convergence came primarily from the choice
of RG transformation.
Indeed,
there is no evidence that the tuned RG transformation
brings the renormalized Hamiltonians closer to the fixed point.
Apparently,
the renormalization trajectory
is such that even though it passes through points
a significant distance from the fixed point,
the convergence of $y_{T1}$
is very good.

The fact that the RG trajectory itself does \emph{not}
converge rapidly,
can be seen in Table \ref{table: nnYT1-majority}.
This table is organized differently than
the tables for the eigenvalue exponents.
Rows correspond to equally-sized renormalized lattices,
with the size indicated by the first column.
The entries are the values of the corresponding correlation functions,
divided by the size of the lattice,
to facilitate comparisons.
It can be seen that although the correlation functions are closer to each other
when they correspond to more renormalization iterations,
they have not converged for the RG iterations down to $4^3$.
A comparison with the entries on the diagonal
in Table \ref{table: YT1}
shows a relatively weak size effect for the eigenvalue exponent values.

%  correlation functions

\begin{table*}%[htb]
\caption{Five correlation functions obtained from
the same simulations as in Table \ref{table: YT1}
and \ref{table: YT2},
with
the parameter $w=0.4314$. 
% and the critical coupling $K_c=0.2216544$.
(a=000-100; b=000-110; c=000-111; d=000-100-010-110 and e=000-100-010-001.) }
\begin{center}
\begin{tabular}{|c|c|c|c|c|c|c|}
\hline
$L^3$  & $\Sigma$ & $256^3$ & $128^3$ & $64^3$ & $32^3$ & $16^3$  \\
%$n$  & $N_{ops}$ & $L=8$ & $L=7$ & $L=6$ & $L=5$ & $L=4$  \\
\hline
$256^3$ & a & 0.330491(1) & & &  & \\
  & b & 0.208951(1) &  & &  &   \\
  & c & 0.163751(2) &  &  &  &    \\
  & d & 0.175667(1) &  &  &  & \\
  & e & 0.115223(1) &  & &  &  \\
\hline
$128^3$ & a & 0.277505(3) &  0.330980(2)  & &  & \\
 & b & 0.193674(4) & 0.209600(3) & &  &    \\
 & c & 0.156671(4) & 0.164487(4) & &  &   \\
 & d & 0.123980(2) & 0.176051(2) &  &  &   \\
 & e & 0.096526(2) & 0.115679(2) &  &  &  \\
\hline
 $64^3$ & a & 0.264324(7) & 0.278867(6) &  0.332281(2) &  &  \\
 & b & 0.189019(8) & 0.195345(8) & 0.211325(3)   &  &    \\
 & c & 0.154033(9) &  0.158523(8) &  0.166440(3) & & \\
 & d & 0.113330(5) &   0.124977(4)  & 0.177075(2)  &  & \\
 & e & 0.091201(5) & 0.097588(4) & 0.116894(2)  & &   \\
\hline
$32^3$ & a & 0.261975(16) &  0.267975(16)  & 0.282485(5)  & 0.335734(4) &    \\
 & b & 0.189760(18) &  0.193421(18) &  0.199785(5)  & 0.215903(5) &    \\
 & c & 0.155992(20) & 0.158896(20) &  0.16350(6)  & 0.171623(5)  &    \\
 & d & 0.111433(12) & 0.115920(11) & 0.127626(4) & 0.179794(3) &    \\
 & e & 0.090799(11) &  0.093920(12) & 0.100414(4) & 0.120119(3) &    \\
\hline
$16^3$ & a & 0.268326(40) &  0.271726(37) & 0.277684(11) & 0.292087(9) &   0.344889(3)   \\
 & b & 0.199029(42) &  0.201440(44) &  0.205131(12) & 0.211566(11)  &  0.228043(3)   \\
 & c & 0.166844(46) &0.168865(48)  &0.171831(13)  & 0.176526(12) &  0.185370(3)  \\
 & d & 0.115740(29) &  0.118288(26) & 0.122836(8) & 0.134672(7) &   0.187019(2)  \\
 & e & 0.095977(28) & 0.097952(28)&  0.101185(8) & 0.107933(7) & 0.128690(2)   \\
\hline
$8^3$ & a & 0.291884(85) & 0.294200(87) &  0.297557(25) &  0.303419(22)  &   0.317542(6)   \\
 & b & 0.228122(96) &   0.229982(95)&  0.232453(29)    & 0.236208(25) & 0.242812(7)   \\
 & c & 0.199280(102) & 0.200925(102)  & 0.203053(32) &  0.206160(26) & 0.211216(7)    \\
 & d & 0.132318(70) & 0.134094(64) &  0.136752(19) &  0.141473(17) &  0.153630(5)  \\
 & e & 0.113640(65) &  0.115173(63) & 0.117304(18) &  0.120794(17) &  0.128180(5) \\
\hline
$4^3$ & a & 0.358390(186) &  0.360473(181) &0.362650(60)   &   0.365979(44) & 0.371601(13)     \\
 & b & 0.307968(197) & 0.309833(214) & 0.311695(66) &  0.314279(50)  &  0.318110(14)  \\
 & c & 0.286644(206) & 0.288429(226) & 0.290147(68) & 0.292477(54) &   0.295778(14)    \\
 & d & 0.183398(168) &  0.185148(152) & 0.187090(51) & 0.190036(37) &  0.195074(11)   \\
 & e & 0.167453(156) &0.169101(150) & 0.170828(51) &   0.173341(38) & 0.177337(10)  \\
\hline
\end{tabular}
\end{center}
\label{table: nnYT1-majority}
\end{table*}

\begin{table*}%[b]
\caption{Estimates of the critical exponents
and the eigenvalue exponents
from several sources.
Values that are marked with an asterisk
are calculated to be consistent with
the published exponents
in the same source.
For the last three columns,
$y_{T1}$ and $y_{H1}$
are obtained for the corresponding values of
$\nu$ and $\eta$.}
\begin{center}
\begin{tabular}{|c|c|c|c|c|c|}
\hline
       & This work  &   Ref.\cite{Bloete_Heringa_Hoogland_Meyer_Smit_1996}  &   Ref.\cite{Hasenbusch_2010}  &   Ref.\cite{Guida_ZinnJustin_1998}  &   Ref.\cite{Guida_ZinnJustin_1998} \\
    &  MCRG&  MCRG&  MC&       $\epsilon$-expansion  &    $d=3$  \\
 \hline
$y_{T1} $     &  1.591(1)    &    1.585(3)    &  1.5872(3)*     &  1.590(63)*    &    1.5862(33)*       \\
$y_{H1}$      &   2.4829(2)   &     2.481(1) &  2.4819(1)*     &   2.4820(25)*    &    2.4833(13)*       \\
$\nu $     &   0.6285(4)*   &   0.6309(2)*  &  0.63002(10)     &   0.6290(25)    &      0.6304(13)     \\
$\eta $     &    0.0342(4)*  &  0.038(2)*   &  0.03627(10)     &     0.0360(50)  &     0.0335(25)      \\
$\beta$      &   0.3250(2)*   &  0.3274(9)*   &    0.32645(10)*   &   0.3257(26)    &      0.3258(14)     \\
$\gamma$      &   1.2356(8)*   &   1.2378(27)*  &   1.2372(4)*    &    1.2355(50)   &   1.2396(13)        \\
$y_{T2} $     &  -0.75(5)   &     &    -0.832(6)*   &    -0.814(19)*    &     -0.799(10)*      \\
$\omega $     &  0.47(3)*    &     &  0.524(4)     &    0.512(12)  &   .504(6)   \\
\hline
\end{tabular}
\end{center}
\label{results}
\end{table*}

%\FloatBarrier

\section{Summary and future work}
\label{section_summary}

The results of our computations
and a comparison with other works
are shown in Table~\ref{results}.
The agreement between the various methods
is generally good,
although some differences exist.
Since we don't have estimates of the systematic errors in our results,
we can't really say what the source of the differences are.

The most reliable of the estimates shown in Table \ref{results}
are those
of Hasenbusch\cite{Hasenbusch_2010}.
This was a very careful Monte Carlo finite-size study
that included many effects of corrections to scaling
to provide limits on the systematic errors.

The largest discrepancies
are between our estimate of $\omega$
and the  estimates
of Hasenbusch\cite{Hasenbusch_2010}
and
Guida and Zinn-Justin\cite{Guida_ZinnJustin_1998}.
Our value, $\omega = 0.47(3)$ is substantially
lower than the others.
This could be partly due to the large fluctuations,
but it could be that
the estimates for
$y_{T2}$ are not yet converged.
Table \ref{table: YT2}
could be easily viewed as indicating that
the absolute values  of
the estimates of $y_{T2}$
are still increasing with the iterations of
the renormalization group.
One possibility for improving the convergence 
is to optimize the RG parameter $w$ separately for the
eigenvalue exponent $y_{T2}$.  
We are currently exploring this possibility.

The most obvious source of systematic error in our calculation 
of the largest eigenvalue exponents
is the uncertainty of the  value of the critical coupling that was used in the 
Monte Carlo simulation.
The usual way to estimate the critical coupling from MCRG
is to use the convergence of the renormalized correlation functions.
Unfortunately, 
it is clear from Table~\ref{table: nnYT1-majority}
that the convergence of the correlation functions 
for the tuned RG transformation
is not sufficient for the purpose.
It might be possible to do the calculation 
with the assistance of an extrapolation of the values,
but it seems more promising to use the convergence 
of the eigenvalues.
Preliminary calculations are very encouraging, 
but more work is needed.

%\FloatBarrier

\makeatletter
\renewcommand\@biblabel[1]{#1. }
\makeatother

\bibliography{Citations_4}

\end{document}